**NUCLEAR ENERGY | ENERGY SUPPLY**

Future energy supply in Germany

# Can nuclear energy contribute to the energy transition?

Axel Kleidon | Harald Lesch

*In the course of the energy transition, energy generation from nuclear power - through nuclear fission and perhaps in the future through nuclear fusion - is often proposed as an alternative or supplement to renewable energy sources. There are already good reasons why electricity generation from nuclear energy is significantly more expensive than other forms of generation, while increasing dryness as a result of climate change is generally calling into question the reliability of thermal power plants. Nuclear energy is therefore unlikely to play a role in a future energy supply that relies on low costs and reliability.*

The current restructuring of our energy system relies on renewable energies, especially photovoltaics, wind energy and biomass. But it also needs storage technologies such as pumped storage power plants, batteries or chemical storage such as green hydrogen to compensate for differences in energy generation and consumption. Nuclear energy and nuclear fusion [1] are also occasionally brought into play as alternatives, especially because they are intended to cover the base load. But how does this fit in with the variable sources of solar and wind? Does it make sense to rely on nuclear energy again?

In this article, we focus on a few key and objective reasons that speak against nuclear energy and nuclear fusion. We primarily look at the costs, i.e. the goal of creating the cheapest possible energy system of the future that is free of fossil fuels. But the availability of water, which is needed to cool thermal power plants, is also critical. Other goals often play a role in nuclear energy, such as military use - as with existing and potential nuclear powers - or the goal of becoming independent of energy imports from abroad in order to avoid being blackmailed - especially in Eastern European countries. In Germany, however, we can achieve the latter goal by using solar and wind power.

## Costs of energy production

The key to determining the contribution of a form of energy generation to society's energy needs is an estimate of its costs, in this case the levelized cost of electricity (LCOE) . To calculate this, the investment costs for construction ($I$), the maintenance costs ($M$) and the costs for fuel ($F$) are added together over the lifetime. We divide this by the expected electricity yield (*Y*) over the lifetime. The equation for the *LCOE* is thus obtained (in units of $/kWh or Euro/kWh):

$$\text{LCOE} = \text{Sum of all costs/Expected yield} = \frac{\sum_{i=1}^{n} (I_i + M_i + F_i)/(1+r)^i}{\sum_{i=1}^{n} Y_i/(1+r)^i} \quad . \quad (1)$$

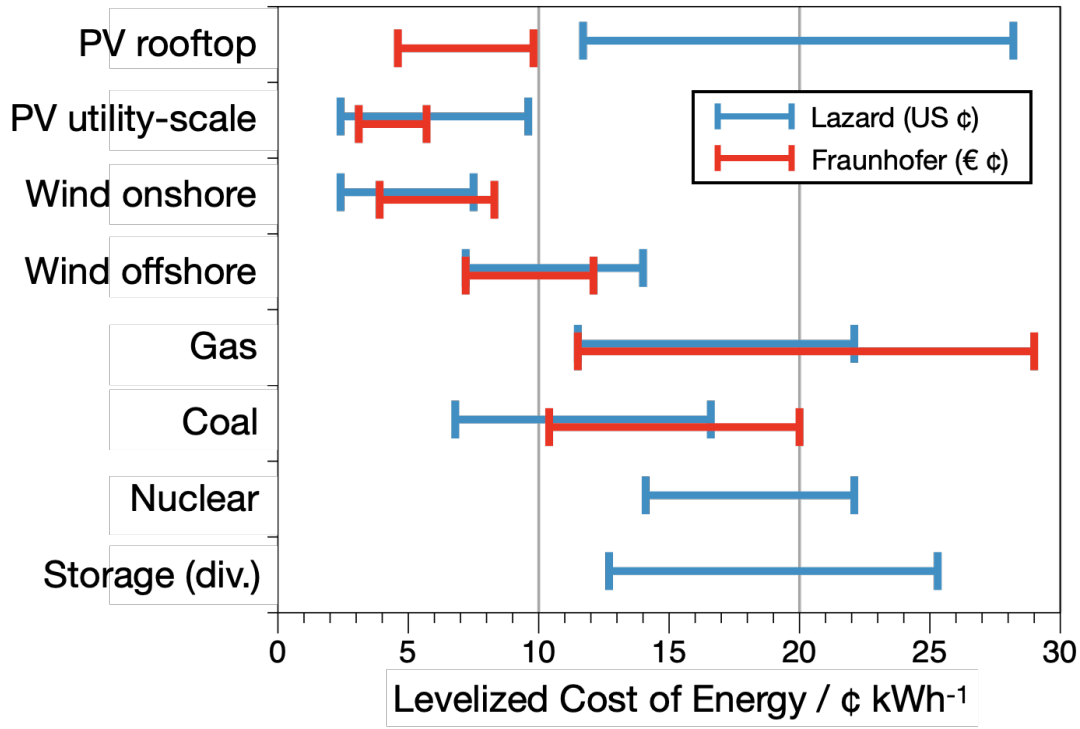

**FIG. 1 ELECTRICITY GENERATION COSTS**
*Comparison of electricity generation costs according to Lazard (blue, international, in US cents per kWh) and Fraunhofer (red, focused on Germany, in Euro cents per kWh). "Storage" includes various forms of storage in the form of mechanical, thermal, or chemical energy. This summary (simplified according to [2] and [3]) excludes subsidies that arise, for example, from the storage of nuclear waste, but are paid for by the state of Germany. The lower costs of rooftop photovoltaics are probably due to experience and lower installation costs in Germany.*

The index *i* describes the year in which the costs are incurred. The totals run over the expected lifetime *n* in years. The expressions in the denominators in the form $(1 + r)^i$ describe the weighting by the discount rate *r*. This economic concept mainly describes the effects of interest rates and inflation. The discount factor can be used to calculate future costs or income back to today. This discount factor is used to weight the costs and benefits over the lifetime. Other costs that arise during operation, such as for $CO_2$ emission certificates from conventional power plants, but also the disposal costs of nuclear power plants, can also be added here. Current values of the production costs for various types of electricity generation are summarised in Figure 1 [2, 3].

Let's take a closer look at the expression for the electricity generation costs. The costs are usually dominated by the investment costs *I* that arise from building a power plant, a wind farm, a solar park or the like. In addition, there are the running costs for operation, maintenance and repair. In the case of conventional power plants, there are also the costs for the fuel - coal, gas, or fissile or fusion material for nuclear or fusion reactors - or for $CO_2$ emissions. With renewable energies such as wind and solar, however, these costs for the fuel or emissions do not arise.

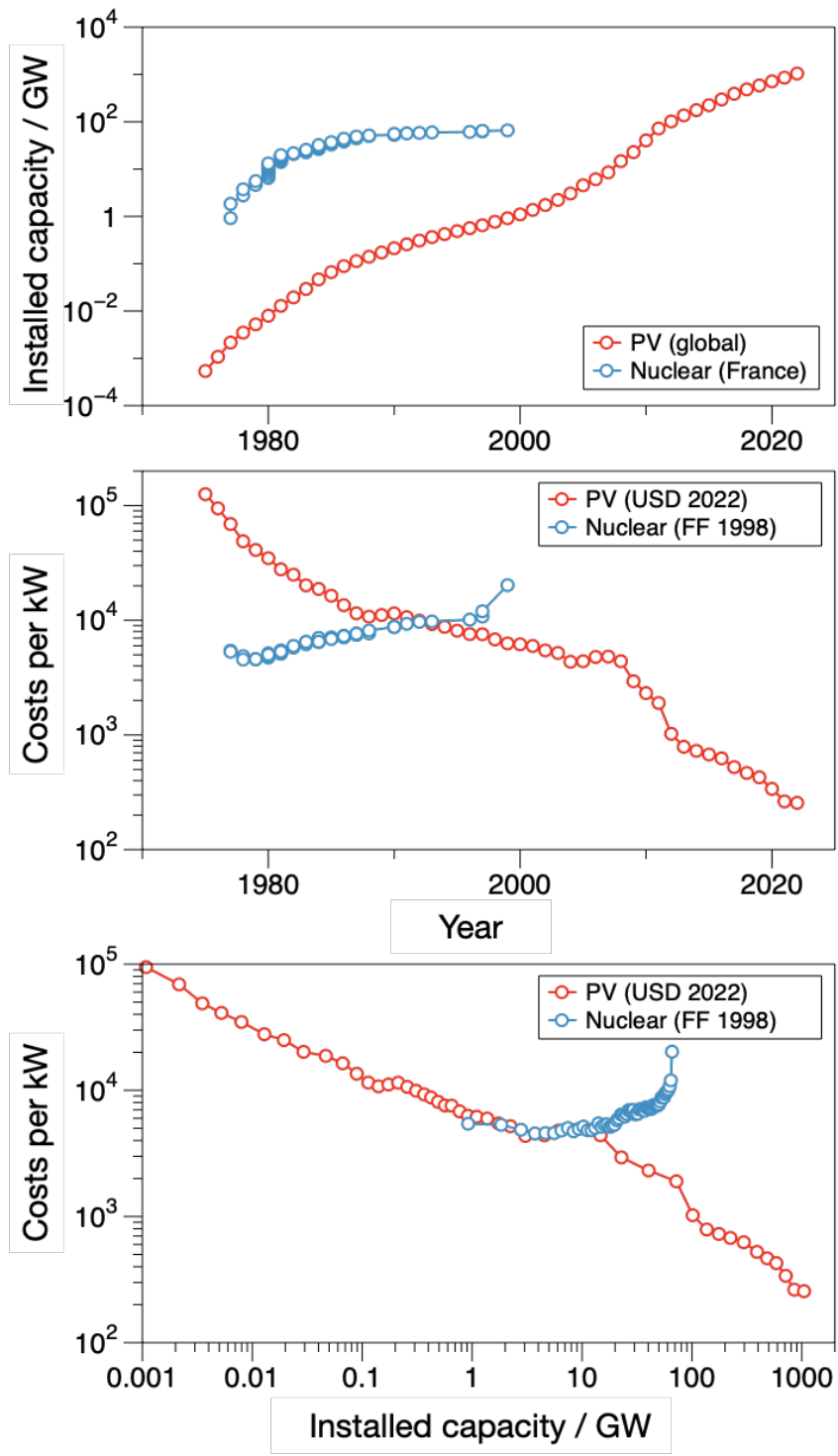

**FIG. 2 COST DEVELOPMENT**
*Cost evolution of photovoltaics (red, in 2022 US dollars) and nuclear power plants (blue, in 1998 French francs) in the past (data: Our World in Data [4], adapted from various sources, [5]) .*

These costs are divided by the electricity yield *Y*, described by the denumerator in equation (1), to determine the average electricity production costs. The electricity yield in turn is described by the capacity of the plant, i.e. how much electricity a plant can generate at most (in megawatts, MW), and the capacity factor, which describes the utilisation. The product of installed capacity and capacity factor then gives the electricity yield. In a thermal power plant, the type of management determines the capacity factor; in renewable energies such as photovoltaics and wind energy, it is mainly determined by the availability of the sun and wind.

We can see that photovoltaics and wind are already cheaper to generate electricity than conventional power plants. We can also see from equation (1) that conventional power plants generate cheap electricity especially when they run continuously, i.e. when they have a high capacity factor. Base load power plants such as nuclear power plants or future fusion power plants are based on this cost model. They are profitable when they continuously generate as much electricity as possible. The high investment costs can then be covered more easily. An extension of the operating time, i.e. a larger *n*, can also contribute to this - then the high investment costs are divided over more electricity yield, and the levelized cost of electricity falls. However, if variable forms of generation contribute more to electricity generation as part of the energy transition, continuously running base load power plants will be needed less and less. They will then become more expensive the more our electricity system is based on renewable energies - an important result that we will come back to.

## Stronger learning curves for simple technologies

First, we look at how the costs of nuclear power plants have developed in the past to see whether we can expect lower costs. This could reduce the contribution of capital costs to the levelised cost of electricity. For nuclear power plants, this contribution dominates, and this will almost certainly be the case for potential fusion power plants as well, as they are even more complex.

In general, the cost of technologies can decrease significantly over time - clearly seen in lasers, LEDs and smartphones. This effect is known as the "economy of scale" or the "learning curve". The learning curve establishes the relationship between cumulative production and the labor costs per unit of product. The more that is produced, the cheaper each unit can be produced.

Of course, this also applies to energy generation. There are two examples that we want to compare in more detail: photovoltaics and nuclear power plants. To do this, we look at how the costs and the capacity produced have developed over the decades, or how the unit costs have changed with the corresponding total capacity produced (Figure 2).

Photovoltaics (PV) is a relatively simple and straightforward technology. Installed capacity is distributed over small units – a PV module typically has a capacity of a few hundred watts. The production of solar modules has increased enormously in recent decades. A global installed capacity of 1 GW was reached in 2000, while in recent years it has increased almost exponentially by more than 100 GW per year. The cost per watt has fallen rapidly. While in the early 1990s a PV capacity of 1 kW still cost around 10 000 US dollars, the cost fell to about 250 US dollars per kW in 2022. The learning curve is accordingly strongly negative: If the installed capacity is doubled, the unit price has decreased by more than 25 % [4].

The combination of the main factors responsible for the rapid price drop in photovoltaic systems is interesting. On the one hand, there are the economies of scale and learning effects already mentioned, but also the technological progress in the manufacturing of solar modules, the production of which has become increasingly automated. In addition, the ongoing technical improvement of the modules has meant that fewer and fewer raw materials are needed for 1 kW$_{peak}$. In addition, the number of suppliers has increased, which also pushes down the price.

A similar combination is difficult to achieve when building nuclear power plants or fusion reactors. In fact, the opposite is observed for nuclear power plants: an analysis [5] of costs in France shows that nuclear power plants there have become significantly more expensive per kW over the 20 years of expansion. While costs initially amounted to around 5000 French francs per kW, they had almost doubled by the 1990s. A look at a few nuclear power plants currently under construction in Europe shows that construction is taking much longer and the costs are much higher than originally planned [6]. The Flamanville nuclear power plant in Normandy with a capacity of 1600 MW was originally supposed to cost 3.3 billion euros. When it goes online in 2024, after an eleven-year delay, the costs will have risen to 23 billion euros [6]. The situation is hardly different for Hinkley Point C in Somerset, England, with a capacity of around 3200 MW. According to current estimates, the nuclear power plant, which is scheduled to be completed in 2026, will cost around 31 billion euros [6] – twice as much as was initially estimated in 2008.

These two very different examples of learning curves can be generalized to a general pattern of learning curves [7]: Simple technologies that can be used flexibly and do not need to be adapted have historically shown strong learning curves - in photovoltaics as well as LEDs, and to a lesser extent in electric cars and wind turbines. Complex technologies, on the other hand, are much less flexible and have to be adapted for each application. In their case, the learning effect is practically non-existent or even negative. Costs are therefore rising - in nuclear power plants as well as in other thermal power plants for coal or geothermal energy.

Conclusion: From the developments and dynamics observed so far, it can be deduced that no significant learning effect is to be expected in nuclear technology. Investment costs are likely to rise. In photovoltaics, on the other hand, costs could continue to fall. Photovoltaics is a comparatively simple technology, it can be used flexibly and shows a strong learning curve. The investment costs per kilowatt in the electricity generation costs could therefore continue to fall, even if it is already the cheapest form of electricity generation (Figure 1). This finding is ultimately reflected in the installed capacity: While around 374 GW of nuclear power plants was installed globally by the end of 2023 and this number is stagnating [6], the figure for photovoltaics was 1055 GW, with a current growth rate of 20 % per year [8].

We do not yet have any experience with nuclear fusion. However, the scientific research institutions that deal with nuclear fusion repeatedly point out the increased complexity of a fusion reactor [9]. It is therefore to be expected that the investment costs per kilowatt will be higher than for nuclear fission.

## Renewables displace the base load

Ideally, the costs of generating electricity are covered by selling the electricity on the electricity market, rather than by government subsidies that can distort the market. With this in mind, let's look at what type of thermal power plants can most cost-effectively fill the gaps when solar and wind power are insufficient.

The electricity market has become more complex and international over the past few decades. Power grids connect producers and consumers across Europe. The variable feed-in of renewable energies leads to differences in supply and demand, which are reflected in fluctuations in the wholesale electricity price. This is shown as an example in Figure 3 for a few days in May 2022 using data from the Bundesnetzagentur (federal agency for infrastructure) for the German electricity market [10]. The daily fluctuations in demand (the "grid load", blue line) are clearly visible; also shown is the total generation (black line) as well as the feed-in from solar (orange) and wind (purple). In times of high feed-in from solar and wind, electricity generation can exceed the grid load. Then the price on the electricity market falls. Examples of this are marked in Figure 3 by the light red sections.

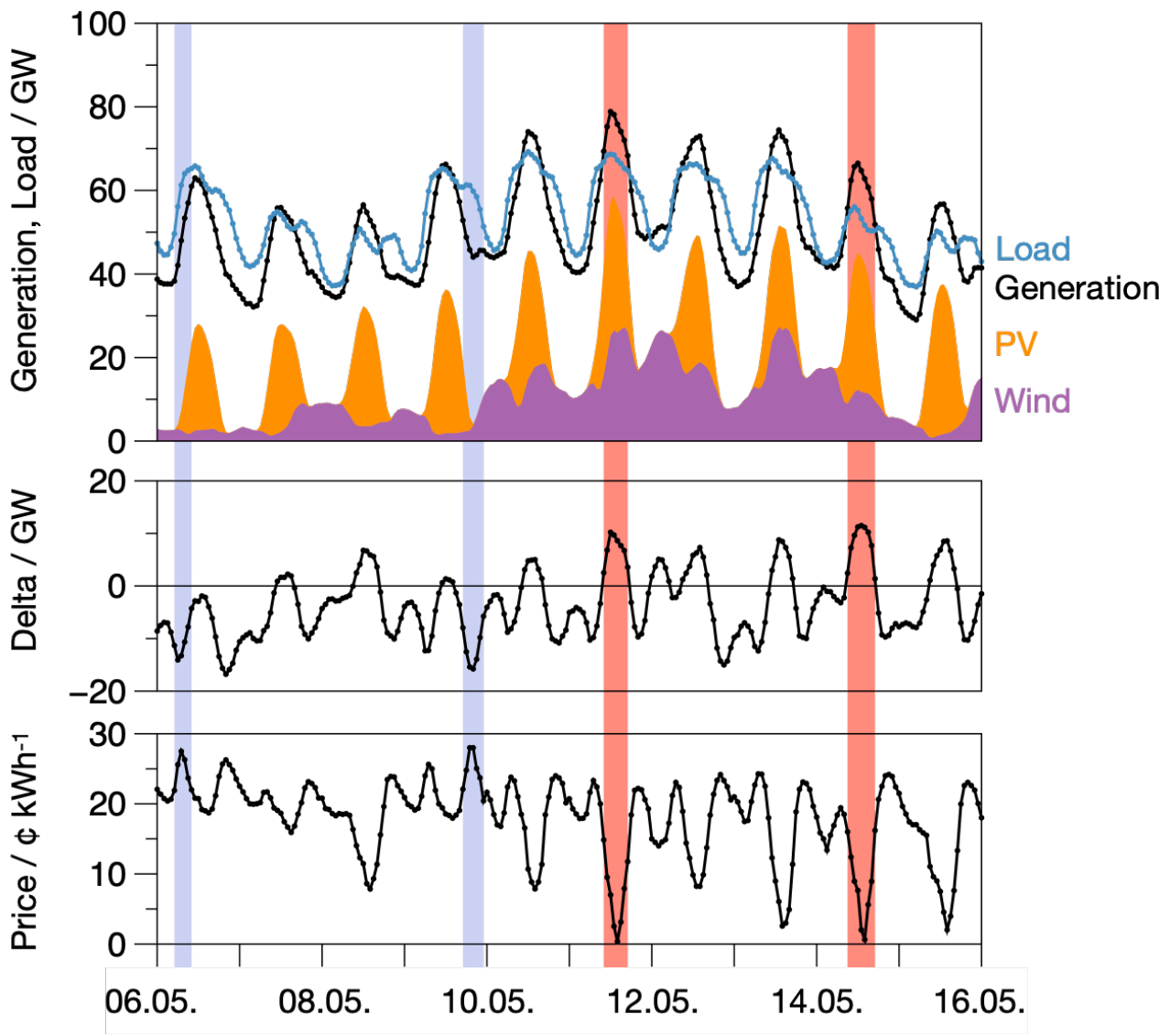

**FIG. 3 ELECTRICITY MARKET**
*Top: Electricity generation and grid load (i.e. consumption) as well as feed-in from solar and wind for a period in May 2022. Middle: Difference between grid load and generation. Bottom: Electricity price on the European Energy Exchange (EEX). Color-coded sections illustrate maxima and minima in the electricity price in relation to variable feed-in (data: [10]) .*

At other times, there is only a small amount of solar and wind power, and they cannot satisfy the grid load. This is when the price of electricity rises (light blue sections in Figure 3). To bridge these times, we need either electricity storage, imports from abroad or feed-in from other forms of generation such as thermal power plants that can generate electricity in a controlled manner.

But which type of thermal power plant can close these gaps most cost-effectively? These gaps do not require base load power plants that feed in electricity more or less continuously throughout the year, but power plants that contribute electricity only now and then. And this increases the levelized cost of electricity because the yield is reduced (i.e. the denominator in equation 1), but the costs remain more or less the same. This is particularly the case with high investment costs,

**TAB. 1 ELECTRICITY GENERATION COSTS**

| Type of power plant | Annual fixed costs (€ kW$^{-1}$ year$^{-1}$) | Variable costs (€ MWh$^{-1}$) |
|---|---|---|
| Nuclear power plant | 420 | 13 |
| Coal-fired power plant | 190 | 19 |
| Gas turbine power plant | 62 | 25 |

Breakdown of annual electricity generation costs for nuclear, coal and gas turbine power plants (based on [3]).

while the costs of power plants with expensive fuel also fall accordingly with reduced running times.

To describe this in more detail, we divide the average annual cost of electricity generation $C_{ann}$ (in Euro kW$^{-1}$ year$^{-1}$) into fixed costs ($C_{fix}$, in Euro kW$^{-1}$ year$^{-1}$) and a variable part ($C_{var}$, in MWh$^{-1}$), which depends on how much electricity the power plant generates:

$$C_{ann} = C_{fix} + C_{var} \cdot f_{cap}. \qquad (2)$$

The fixed costs are mainly determined by the investment costs, i.e. $I$ in equation (1), the variable costs largely by the fuel costs, $F$ in equation (1). This includes the capacity factor $f_{cap}$ (expressed in full load hours per year), which describes the utilization of the power plant. For base load power plants, the capacity factor is close to 100% or 8760 h year$^{-1}$.

Different types of thermal power plants differ in terms of their fixed costs and variable share. Estimates for these shares are given in Table 1 summarised from the same data source [3] that was also used in Figure 1. The costs are shown graphically in Figure 4 (top). Gas-fired power plants require little investment, so they are cheap to build, coal-fired power plants are more expensive, but nuclear power plants are by far the most expensive. The variable contribution also differs between the different types of power plants, but only to a lesser extent. However, this can change accordingly if $CO_2$ prices increase, for example.

From this breakdown, we can clearly see the effect of reduced feed-in of thermal power plants (Figure 4, middle). The production costs per amount of electricity generated increase the lower the utilization of the power plant. This is particularly the case with nuclear power plants, as the investment costs are highest here, while the effect is significantly smaller for gas-fired power plants.

What happens now when the feed-in from solar and wind increases? To illustrate this, we first look at the electricity consumption broken down over time. We get this from the frequency distribution of the grid load, the so-called load curve. It is shown in Figure 4 bottom for the year 2022, determined from hourly data from the Bundesnetzagentur [10]. It can be interpreted as follows: The maximum value of the grid load was 78.7 GW, it occurred in one hour of 2022 and represents only 0.01 % of the year. The lowest value was 34.4 GW, which means that in 2022 the grid load was always at least 34.4 GW. Without renewable energies, we can directly see from this distribution how many base load power plants we need - namely, enough for a guaranteed output of 34.4 GW.

If the sun and wind now contribute to electricity generation, the remaining load that still needs to be generated changes. This is the so-called residual load, i.e. the grid load reduced by the feed-in from the sun and wind. This load, which has to be satisfied by thermal power plants or other forms of electricity storage, therefore drops to lower values: the maximum load drops to 70.9 GW, the lowest value is now 3.1 GW, which is significantly less than without the feed-in from the sun and wind. The need for base load power plants has therefore already been significantly reduced.

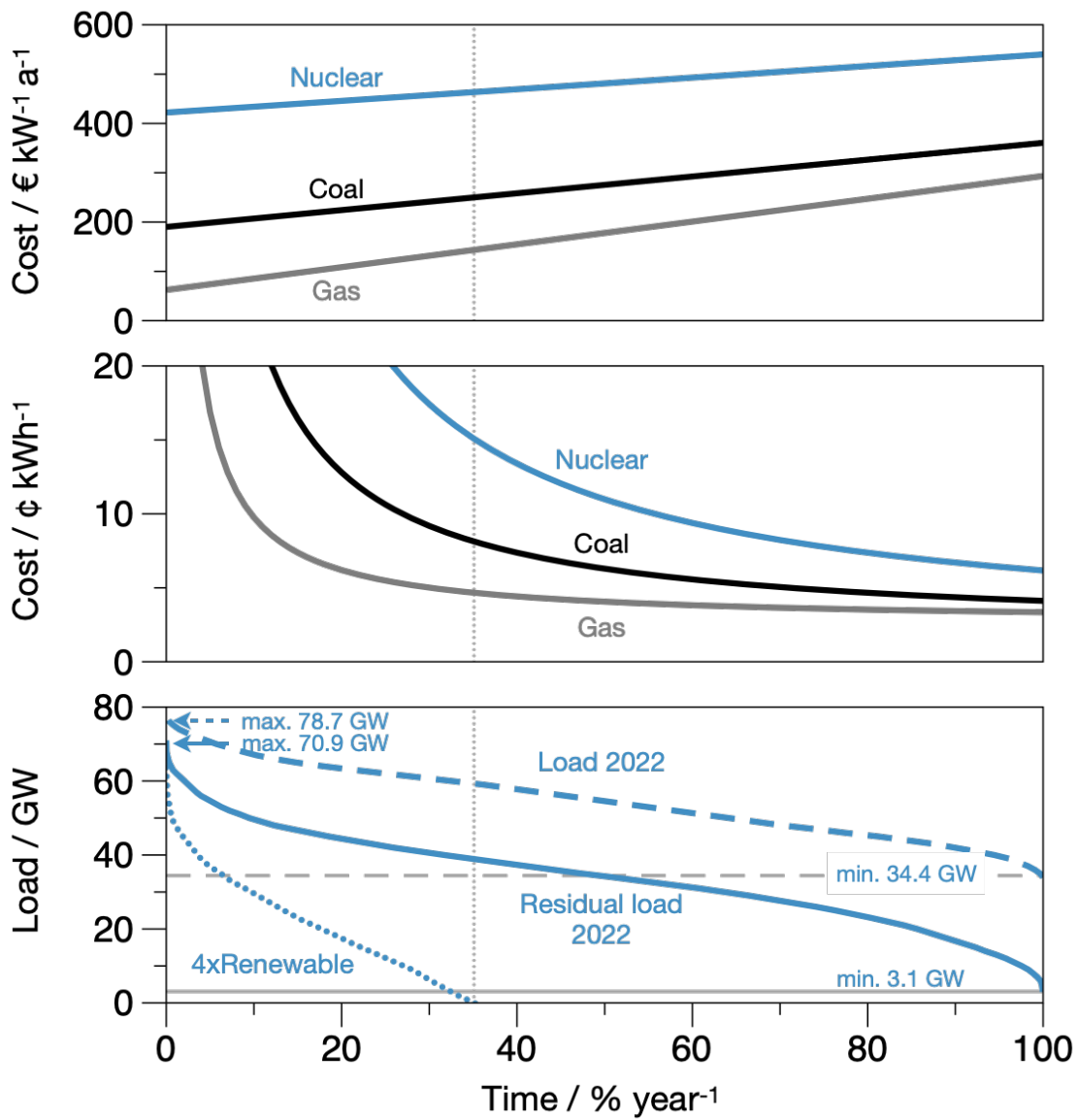

**FIG. 4 ELECTRICITY GENERATION COSTS**
*Top: Annual electricity generation costs of gas, coal and nuclear power plants as a function of their capacity utilization, shown as a percentage of the year in which they generate electricity. Middle: Annual electricity generation costs as a function of capacity utilization. Bottom: Distribution of grid load (dashed) and residual load (solid, i.e. grid load reduced by the feed-in from solar and wind) for the year 2022. Also shown is a scenario (dotted) in which the feed-in from renewables was quadrupled (based on [3] and [10]).*

With the expansion of renewable energies, the residual load continues to fall. This is shown in Figure 4 by the dotted line. It represents a hypothetical scenario in which the feed-in from solar and wind is increased by a factor of 4, whereby the increase in electricity demand due to modernization through electrification - such as heat pumps and e-mobility - is not taken into account here. This roughly corresponds to the planned expansion by 2050. We can see that the need for base load has completely disappeared. Thermal power plants will then only be needed for about a third of the year to generate electricity. The capacity factor thus drops to 35 %. They – or electricity storage – are still needed to fill the gaps, also because the peak load is still 64.2 GW.

This then leads to electricity generation from thermal power plants becoming more expensive because they are no longer operated all year round, meaning their capacity factor is falling. However, the increase in price is comparatively small for gas-fired power plants because the investment costs are low. $CO_2$ emission certificates will become more expensive in the foreseeable future, thereby increasing fuel costs. But these power plants will also be operated with green hydrogen in the future, i.e. without $CO_2$ emissions, which is extracted or imported from times of excess electricity. Electricity storage, international electricity trading or changes in electricity usage behavior by consumers can also reduce the demand for these power plants. In the case of nuclear power plants with their high investment costs, on the other hand, the amount of electricity generated becomes considerably more expensive. For this reason, typical scenarios of the energy transition [11] rely on gas-fired power plants as flexible and cheapest electricity generators that fill the gaps left by solar and wind.

With the continued expansion of renewable energies, thermal power plants and storage facilities will still be needed to fill generation gaps, even if these gaps are getting smaller. Power plants with low investment costs can do this much more cheaply, even if they have comparatively high fuel costs. The investment costs for nuclear power plants are very high, which is reflected heavily in their electricity price when capacity utilization is low.

## Less cooling water due to climate change?

Another aspect that may limit nuclear energy and thermal power plants in general in the future is the increasing drought in Germany. After all, such power plants need a lot of cooling to maintain a temperature gradient in the turbine circuit.

This cooling is achieved either by evaporating water in cooling towers or by heating river or sea water. In order to generate an output of 1000 MW of electricity, a thermal power plant with an efficiency of 40 % needs 2500 MW of heat. The difference of 1500 MW is the waste heat that must be removed by cooling. With a heat of vaporization of 2.5 MJ per kilogram of water, this can be accomplished by evaporating 0.6 tons of water per second (0.6 $m^3$/s or 18.9 x $10^6$ $m^3$/year), or by heating 36 $m^3$/s of river water (1.1 x $10^9$ $m^3$/year) by 10 K (with a specific heat capacity of 4180 J $kg^{-1}$ $K^{-1}$). In Germany, the grid load in 2022 was about 500 TWh, or 57 GW. If this electricity had been generated entirely by thermal power plants, this would correspond to 1.1 x $10^9$ $m^3$ $year^{-1}$ of evaporated water or 64.5 x $10^9$ $m^3$ $year^{-1}$ of heated river water. For comparison: the evaporation rate of water per year corresponds to 2.3% of Lake Constance's volume of 48 $km^3$, and the river water consumption per year would correspond to 30% more than the entire volume of Lake Constance.

To relate these water flows to the availability of freshwater in Germany, we look at the so-called potential water supply (Figure 5 top). Precipitation and the inflow of the Rhine, Elbe and Oder from our neighboring countries provide us with freshwater, but some of it evaporates through forests, meadows and fields. Balancing these flows gives us a measure of the rate at which we could use freshwater sustainably. In the period 1991-2020 it was about 176 x $10^9$ $m^3$ $year^{-1}$ [12]. Almost 40 %, or 69 x $10^9$ $m^3$ $year^{-1}$, are contributed by the inflow from neighbouring countries, the rest of 105 x $10^9$ $m^3$ $year^{-1}$ comes from the excess of precipitation minus evaporation.

Of the water supply, about 20 % is used for human consumption for drinking water, industry and energy production (Figure 5 middle), of which about 44 %, i.e. 8.8 x $10^9$ $m^3$ $year^{-1}$, for cooling thermal power plants. With the expansion of renewable energies over the last decades, water consumption for thermal power plants has fallen significantly (Figure 5 bottom): In the early 1990s, our electricity was mainly generated by thermal power plants, and accordingly more than three times as much water was needed for cooling [12]. If more electricity is to be generated from nuclear energy, more cooling water is required and, accordingly, greater use of water resources.

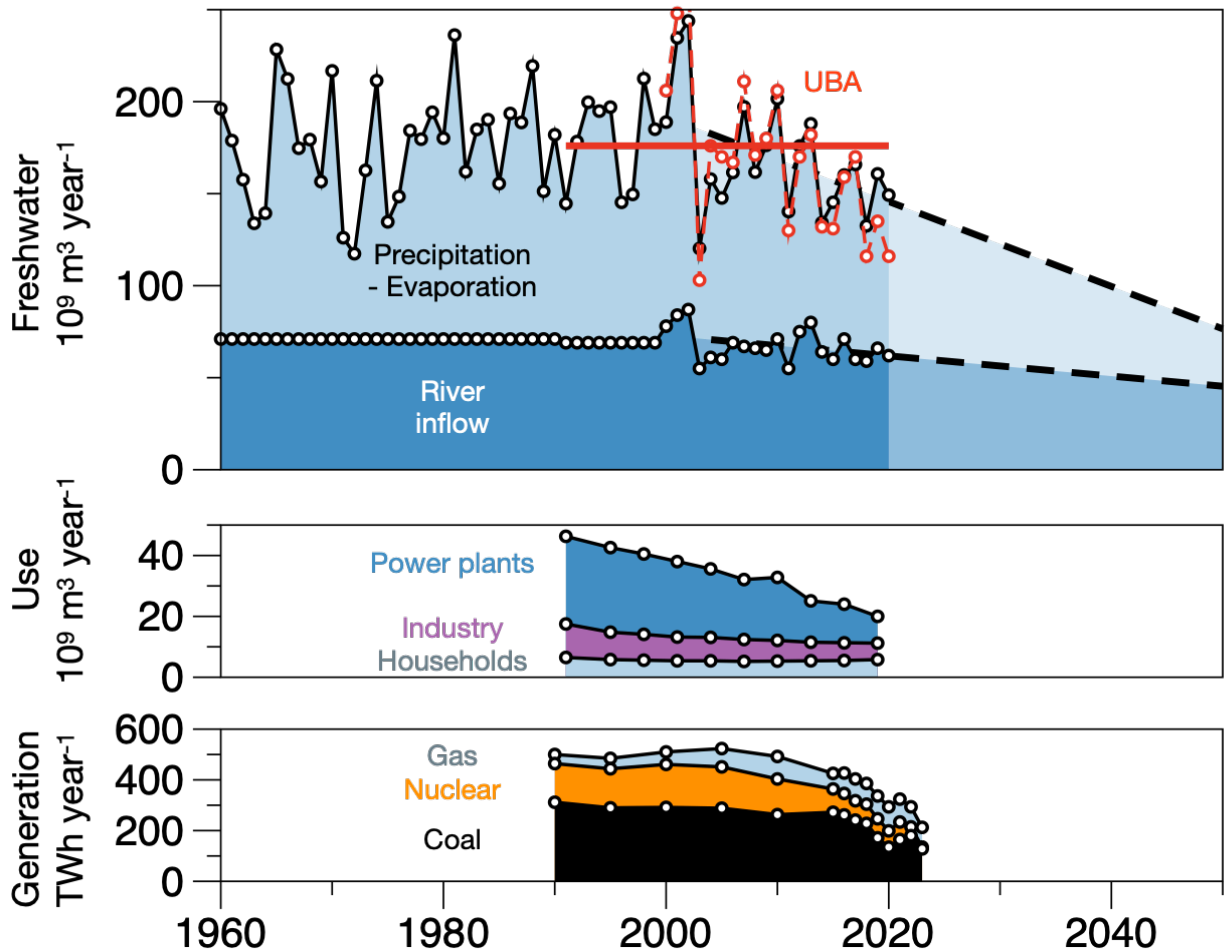

**FIG. 5 WATER AVAILABILITY IN GERMANY**
*The potential water supply of Germany (top) and the contributions of inflow and the surplus from precipitation and evaporation, calculated from data from the German weather service (DWD) and the federal environmental agency of Germany (UBA) (red), as well as use of freshwater (middle). Bottom: electricity generation from thermal power plants is shown for comparison (sources: [12, 13]) .*

However, the potential water supply has decreased significantly over the last 20 years (see dashed lines in Figure 5 top). This is mainly due to reduced precipitation, although the reason for this is not fully understood. It could simply be decadal climate variability, but it could also be a trend in climate change. What has clearly increased, however, is the capacity to evaporate water [13]. This is described by the concept of potential evaporation, i.e. the rate at which water could evaporate if its availability is not restricted.

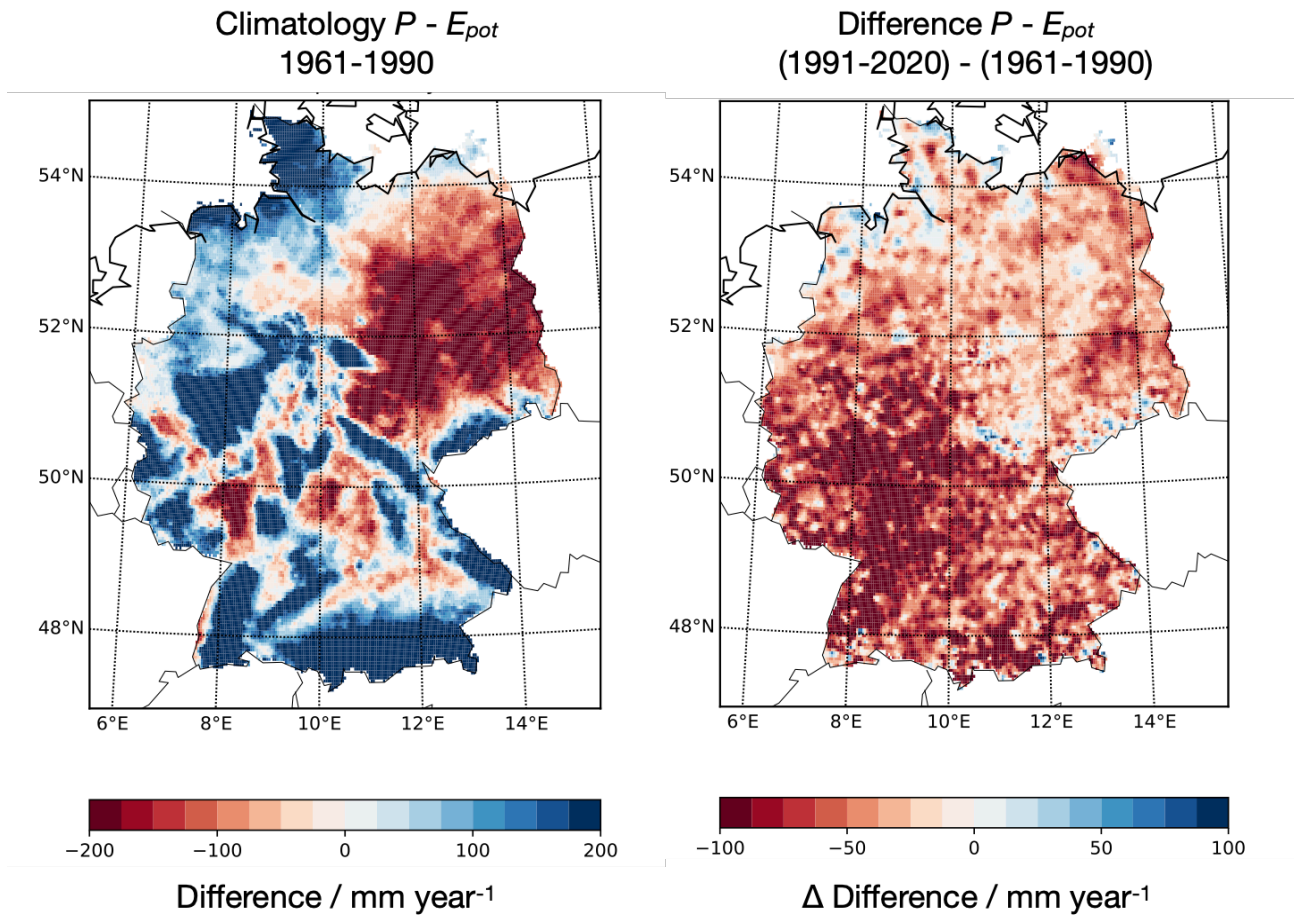

**FIG. 6 DROUGHTS IN GERMANY**
*Climatological difference between precipitation and potential evaporation for the climatological reference period 1961-1990 and the change over the period 1991-2020 (Source: [13]).*

Real evaporation can be close to potential evaporation in forests, especially if they reach enough soil water via their root systems to compensate for seasonal differences between precipitation and potential evaporation. In agricultural land, on the other hand, it is typically well below the potential, especially in summer after the crops have been harvested and fields are left fallow. The difference between precipitation and potential evaporation nevertheless provides a measure of climatological water availability that does not depend on land use, but only on natural availability of energy and precipitation.

This difference is shown in Figure 6 for the climatological reference period from 1961-1990 and the difference for the following 30 years (1991-2020). The east of Germany is dry, so there is a climatological water deficit: potential evaporation is higher than precipitation. The river water comes either through the inflow to eastern Germany or from the low mountain ranges and Alpine foothills, where precipitation is significantly higher than potential evaporation.

However, potential evaporation has clearly increased over the last 30 years as a result of climate change (Figure 6). There are simple physical reasons for this: firstly, the last five years have been on average 2 °C warmer than the climatological reference period from 1961 to 1990 [13]. At warmer temperatures, air can hold more water vapor while its heat storage capacity remains the same. When sunlight is absorbed and the surface is warmed, the air near the ground is more humidified at warmer temperatures, so more evaporation occurs. Secondly, Germany has become sunnier over the last 30 years. This has contributed about half of the observed warming, thus increasing potential evaporation. This means that at least part of the trend of decreasing water supply in recent years can be explained by climate change.

Thermal power plants need a lot of cooling water, although climatological trends are currently reducing the water supply. This effect is further amplified seasonally, as potential evaporation increases particularly in the warm summer months. When more water is then needed for irrigation in agriculture, the availability of cooling water is no longer guaranteed. This is not the case with electricity generation from solar and wind, as they do not require cooling water.

## Conclusions

For Germany's future energy supply, nuclear power plants are likely to be simply too expensive if the goal is to provide cheap and sustainable energy. Solar and wind power are already the cheapest electricity generation technologies today. Their further expansion means that thermal power plants will run for ever shorter periods of time, meaning that base load power plants will no longer be needed.

The supply gaps - the dark doldrums - can then be covered by thermal power plants with low investment costs such as gas turbine power plants, but also by battery storage or imports. Natural gas can already reduce $CO_2$ emissions today because it releases less $CO_2$ per energy than coal, and it can later be replaced by green hydrogen as an energy source. This need to "fill the gap" will remain for some time, and flexible, inexpensive thermal power plants are one way to achieve this.

## Summary

*Compared to renewable energies, nuclear power plants are significantly more expensive in terms of electricity generation costs. In addition, nuclear power plants have become increasingly more expensive per kilowatt of output over the past few decades, while renewables have become increasingly cheaper. In addition, the variable feed-in from solar and wind energy reduces the remaining grid load that still has to be covered. The remaining gaps in electricity generation can be easily filled by gas-fired power plants. These have significantly lower investment costs than nuclear power plants. If the current trends in water availability as a result of climate change continue, the lack of cooling water could also limit the contribution of thermal power plants to electricity generation, especially in the dry summer months. Nuclear energy in particular is therefore facing two problems: it is very expensive and no longer offers energy security in future dry summers.*

## Keywords

## The authors

Axel Kleidon studied physics and meteorology at the University of Hamburg and Purdue University, Indiana, USA. After completing his doctorate at the Max Planck Institute for Meteorology, he conducted research at Stanford University in California and at the University of Maryland. Since 2006, he heads the "Biospheric Theory and Modeling" group at the Max Planck Institute for Biogeochemistry in Jena. His research interests range from the thermodynamics of the Earth system to the natural limits of renewable energy sources.

Harald Lesch studied physics and philosophy at the universities of Giessen and Bonn and received his doctorate from the University of Bonn. He then conducted research in Heidelberg and Toronto. After his habilitation in 1994 at the University of Bonn, he became full professor for theoretical astrophysics at the LMU in 1995 and has been a lecturer for natural philosophy at the Munich School of Philosophy since 2002. He works on complex cosmic and terrestrial systems up to the natural limits of technological societies.

## Contact

Dr. Axel Kleidon, Max Planck Institute for Biogeochemistry, P.O. Box 10 01 64, 07701 Jena, Germany. akleidon@bgc-jena.mpg.de